\documentclass[journal]{IEEEtran}
\usepackage{tipa}
\usepackage{bbm}
\usepackage{mathrsfs}
\usepackage{cite,url,subfigure,epsfig,graphicx}
\usepackage{amssymb,amsmath}
\usepackage{indentfirst}
\usepackage{algorithmic}
\usepackage{algorithm}
\usepackage{pdfpages}
\usepackage{times,verbatim,amsfonts,amsmath,color}
\usepackage{pifont}
\usepackage{multirow}
\usepackage{booktabs}
\usepackage{adjustbox}
\usepackage{hyperref}
\usepackage{threeparttable}
\usepackage{pifont}

\usepackage{booktabs}  
\usepackage{array}    
\usepackage{makecell} 
\usepackage[edges]{forest}
\usepackage{xcolor}

\IEEEoverridecommandlockouts
\makeatletter
\def\ps@headings{\def\@oddhead{\mbox{}\scriptsize\rightmark \hfil \thepage}\def\@evenhead{\scriptsize\thepage \hfil \leftmark\mbox{}}\def\@oddfoot{}\def\@evenfoot{}}
\makeatother 

\newcommand{\tabincell}[2]{\begin{tabular}{@{}#1@{}}#2\end{tabular}}

\begin{document}
\title{Security of Internet of Agents: Attacks and Countermeasures}
\author{Yuntao~Wang, Yanghe~Pan, Shaolong~Guo, and Zhou~Su
\thanks{Y.~Wang, Y.~Pan, S.~Guo, and Z.~Su are with the School of Cyber Science and Engineering, Xi'an Jiaotong University, Xi'an, China. \textit{(Corresponding author: Zhou~Su)}}
}

\maketitle

\begin{abstract}
With the rise of large language and vision-language models, AI agents have evolved into autonomous, interactive systems capable of perception, reasoning, and decision-making. As they proliferate across virtual and physical domains, the Internet of Agents (IoA) has emerged as a key infrastructure for enabling scalable and secure coordination among heterogeneous agents. This survey offers a comprehensive examination of the security and privacy landscape in IoA systems. We begin by outlining the IoA architecture and its distinct vulnerabilities compared to traditional networks, focusing on four critical aspects: identity authentication threats, cross-agent trust issues, embodied security, and privacy risks. We then review existing and emerging defense mechanisms and highlight persistent challenges. Finally, we identify open research directions to advance the development of resilient and privacy-preserving IoA ecosystems.
\end{abstract}


\IEEEpeerreviewmaketitle
\section{Introduction}
The advent of large language models (LLMs) and vision-language models has transformed AI agents into fully autonomous and interactive entities capable of independent perception, reasoning, and action \cite{wang2024large}. These AI agents (or called agentic AI) \cite{zhao2024see}, ranging from digital assistants to unmanned aerial vehicles (UAVs) and service robots, operate across virtual and physical domains, driving unprecedented demands for an infrastructure that natively supports agent-to-agent (A2A) interactions. Gartner projects that \cite{GartnerAgent}, by 2028, AI agents will autonomously manage at least 15 \% of routine daily tasks, while roughly one-third of enterprise applications will embed agent-based intelligence.
The Internet of agents (IoA) \cite{chen2025ioa,aminiranjbar2024dawn}, also referred to as the agentic web, has emerged to meet this need, offering an agent-centric fabric that supports on-the-fly agent discovery, goal-driven communication, and coordinated task execution at scale. Unlike the traditional Internet, IoA communications focus on machine-readable objects (e.g., model checkpoints, encrypted tokens, and latent embeddings), and agent protocols emphasize semantic negotiation and adaptive orchestration. By pooling distributed inference and shared sensing capacities, IoA extends advanced AI capabilities to resource-constrained devices and establishes new connectivity patterns across heterogeneous agent ecosystems.


As agents process and exchange large volumes of personal and sensitive data, ranging from user profiles and behavioral histories to real-time sensor feeds, they become prime targets for sophisticated cyber adversaries. Malicious actors may exploit agent forgery to impersonate legitimate agents and infiltrate sensitive workflows \cite{jiang2024can}, or employ intent deception to subtly manipulate decision-making logic and contaminate collaborative outcomes \cite{10669201}. Colluding agents can coordinate to distort shared insights or hijack consensus mechanisms \cite{NEURIPS2024_861f7dad}, undermining the integrity of distributed reasoning processes. Meanwhile, adversarial inputs crafted to trigger misclassification \cite{khan2025agents}, contextual backdoors that activate under specific environmental cues \cite{jiao2025can}, and hallucination cascades that propagate spurious outputs across agent networks \cite{zhang2023language} can cause systemic breakdowns in multi-agent coordination.
With the ongoing evolution of IoA, innovative built-in security and privacy preservation mechanisms are essential to realize trustworthy, secure, and privacy-preserving deployments of large-scale IoA ecosystems and unleash the transformative power of future autonomous AI ecosystems.


Recent research in LLM-based agents has attracted considerable attention across both academia and industry.
Das \textit{et al.} \cite{Das2025Security} provide an in-depth survey of security and privacy vulnerabilities in LLMs, evaluating domain-specific risks and defenses across transportation, education, and healthcare. He \textit{et al.} \cite{he2024emerged} categorize emerging threats to LLM-driven agents, illustrates their real-world impacts, and reviews prevailing mitigation techniques. Wang \textit{et al.} \cite{wang2024unique} examine attack vectors at five critical stages: pre-training, fine-tuning, retrieval-augmented generation (RAG), deployment, and in-agent operation, as well as tailored countermeasures. Gan \textit{et al.} \cite{gan2024navigating} introduce a two-axis taxonomy of security challenges in LLM agents by threat source and impact, and analyze representative agent implementations as case studies. Li \textit{et al.} \cite{li2024personal} outline the architecture and optimization workflows of personal LLM agents, highlighting associated security and privacy concerns. Zhang \textit{et al.} \emph{et al.} \cite{zhang2021physical} focus on safety and security issues in agent systems by examining physical faults and cyber attacks including denial-of-service (DoS) and deception attacks. They also discuss countermeasures including fault estimation, detection, diagnosis, fault-tolerant control, and secure cyber attack management.
Deng \textit{et al.} \cite{Deng2025AIagentSecurity} identify four key vulnerability domains in software-form AI agents: complex multi-step inputs, opaque internal executions, environmental variability, and untrusted external interactions. Wang \textit{et al.} \cite{wang2025comprehensive} systematically survey the full-stack safety threats for LLMs and agents by considering life-cycle LLM risks during LLM training, deployment, and commercialization.
Existing surveys mainly focus on security and privacy threats and defense at the single LLM agent level. In contrast, this survey investigates the networking aspects of large model agents within the Internet of agents (IoA), exposing its unique threat landscape, defense techniques, and research gaps. Table~\ref{contribution} summarizes our survey's contributions with previous survey efforts.

\begin{table}[!t]
   \centering \setlength{\abovecaptionskip}{0cm}
    \caption{A Comparison of Our Survey with Relevant Surveys}\label{contribution}
    \resizebox{1.01\linewidth}{!}{
        \begin{tabular}{c|c|l}
        \toprule
        \textbf{Year} & \textbf{Ref.} & \textbf{Contribution} \\ \hline
        2021 & \cite{zhang2021physical} &
          \tabincell{l}{Survey on cyber-physical threats and security in agent systems,\\
                        covering physical faults, DoS, deception attacks, and defenses.} \\ \hline

        2024 & \cite{he2024emerged} &
          \tabincell{l}{Categorize emerging threats to LLM-driven agents, illustrate \\real-world impacts, and review mitigation techniques.} \\ \hline

        2024 & \cite{wang2024unique} &
          \tabincell{l}{Examine attack vectors across five stages (pre-training, fine-tuning,\\
                        RAG, deployment, in-agent operation) and review tailored defenses.} \\ \hline

        2024 & \cite{gan2024navigating} &
          \tabincell{l}{Introduce a two-axis taxonomy of LLM-agent security challenges \\
                        by threat source and impact with case-study analyses.} \\ \hline

        2024 & \cite{li2024personal} &
          \tabincell{l}{Outline architectures and optimization workflows of personal LLM\\
                        agents, highlight security and privacy concerns.} \\ \hline

        2025 & \cite{Das2025Security} &
          \tabincell{l}{Provide an in-depth survey of security and privacy vulnerabilities\\
                         in LLMs, assess domain-specific risks and defenses in \\
                        transportation, education, and healthcare.} \\ \hline

        2025 & \cite{Deng2025AIagentSecurity} &
          \tabincell{l}{Identify four key vulnerability types of agents: multi-step inputs,\\
                        opaque execution, environmental variability, untrusted interactions.} \\ \hline

        2025 & \cite{wang2025comprehensive} &
          \tabincell{l}{Survey full-stack safety of LLMs and agents from \\LLM training, deployment, and commercialization.} \\ \hline

        Now  & \textbf{Ours} &
          \tabincell{l}{Comprehensive survey of security/privacy threats in IoA,\\
                    emerging/potential countermeasures, and open research challenges.} \\
        \bottomrule
        \end{tabular}}
\end{table}

This paper provides a systematic review of security and privacy in the IoA, charting its threat landscape, defense mechanisms, and research challenges to support large-scale deployment. Our objectives are twofold: (1) assess the landscape of security and privacy risks inherent to agent-centric IoA systems, including the scope and impact of security/privacy challenges; and (2) propose effective strategies and solutions to mitigate these threats, ultimately guiding the robust and secure deployment of IoA in various intelligent applications.
Our key contributions include:
\begin{itemize}
    \item \textit{Taxonomy of IoA Threats and Defenses.} We offer an in-depth review of the emerging security and privacy vulnerabilities in IoA across four aspects: agent identity authentication, cross-agent trust, embodied agent security, and privacy threats. For each category, we survey state-of-the-art and potential countermeasures and practical challenges in IoA.
    \item \textit{Open Challenges in IoA.} We highlight critical gaps in current IoA research and outline a research roadmap to guide the development of resilient, trustworthy, privacy-preserving, and ethical IoA ecosystems.
\end{itemize}

\subsection{Paper Organization}\label{subsec:organization}
The remainder of this paper is organized as follows. In Section~\ref{sec:OVERVIEWandLandscape}, we provide an overview of IoA and its security and privacy landscape. Next, we investigate the security and privacy aspects, including agent identity authentication in Section~\ref{subsec:threat1}, cross-agent trust issues in Section~\ref{subsec:threat2}, embodied agent security in Section~\ref{subsec:threat3}, and privacy threats in Section~\ref{subsec:threat4}. Lastly, Section~\ref{sec:FUTUREWORK} outlines future research trends in IoA domain. Table~\ref{table:abbr} summarizes key acronyms used in this survey.

\begin{table}[!t]
   \centering \setlength{\abovecaptionskip}{0cm}
	\caption{Summary of Key Abbreviations in Alphabetical Order}\label{table:abbr}
    \resizebox{1.01\linewidth}{!}{
		\begin{tabular}{ll|ll}
			\toprule
			\textbf{Abbr.}  &\textbf{Definition}            & \textbf{Abbr.} &\textbf{Definition}           \\
			\midrule
A2A   & Agent-to-Agent       & AI  & Artificial Intelligence \\
ANP   & Agent Network Protocol & BLOS  & Beyond-Line-of-Sight      \\
DID &Decentralized IDentifier  &DoS   &Denial-of-Service \\ EMI &ElectroMagnetic Interference &IoA &Internet of Agents \\
IMU &Inertial Measurement Unit &LLM  &Large Language Model    \\
MEMS  &Micro-Electro Mechanical System  &MCP  &Model Context Protocol \\
mmWave &millimeter-Wave &P2P &Peer-to-Peer  \\
PII  &Personally Identifiable Information  &RL &Reinforcement Learning \\
RAG  &Retrieval-Augmented Generation &UAV &Unmanned Aerial Vehicle \\
			\bottomrule
		\end{tabular} }
\end{table}

\section{Overview of Internet of Agents and Its Security and Privacy Landscape}\label{sec:OVERVIEWandLandscape}
In this section, we first introduce the key concept of the IoA with its distinctive paradigm and overview representative agent communication protocols. We then discuss the security and privacy landscape in IoA, highlighting how traditional network risks evolve in agentic environments and identifying novel threat vectors.

\subsection{Overview of IoA}\label{subsec:Overview}
The IoA is an emerging infrastructure in which autonomous software and embodied agents, ranging from personal LLM assistants to industrial robots, seamlessly interconnect, discover one another, and collaborate to accomplish complex tasks \cite{chen2025ioa,aminiranjbar2024dawn}. Unlike prior Web generations centered on human navigation, the IoA is agent-centric:
\begin{itemize}
    \item \textit{Agents as the new entry point.} Rather than humans directly navigating the Internet via personal computers in Web 1.0 or mobile devices in Web 2.0, autonomous agents in IoA navigate and interact with the digital world. A specialized \textit{super personal assistant agent} will act on behalf of its human owner, negotiating on the user’s behalf through personalized UIs while interacting with other agents via APIs or protocols on the backend \cite{chen2024persona}. Simultaneously, numerous non-user-facing agents, representing banks, schools, restaurants, and so on, will interact indirectly with personal assistants to deliver tailored services.
    \item \textit{Flat \& self-organizing agent collaboration networks.} By autonomously organizing and negotiating, all agents regardless of corporate or platform affiliations can establish efficient, task-driven collaboration networks, to dynamically allocate resources and expertise, via self-organization and self-negotiation. Ultimately, the agentic web will evolve into a flatter, more decentralized digital ecosystem \cite{chen2025ioa}.
    \item \textit{Shared intelligence and capability sharing.}
    Beyond basic connectivity, the IoA enables agents to share inference workloads and sensing data at scale. Resource-constrained agents can offload inference tasks, access high-end models, and leverage collective knowledge, achieving on-demand large-model-as-a-service (LMaaS). Besides, shared sensing capabilities in IoA empowers agents particularly embodied ones with beyond-line-of-sight (BLoS) perception.
\end{itemize}

\subsection{Representative Agent Communication Protocols}\label{subsec:Protocol}
Standardized agent communication protocols are key enablers for the IoA to facilitate seamless interaction, coordination, and secure interoperability among heterogeneous agents \cite{MCP,A2A,ANP,Agora}. Recently, the following representative protocols are developed to support structured messaging, identity verification, and tool integration within distributed agent ecosystems.

\textit{1) Model Context Protocol (MCP) \cite{MCP}.} Anthropic's MCP provides a modular interface standard to enable real-time interactions between large models and external tools, services, or data sources. By abstracting tool execution and contextual data access behind a unified protocol, MCP decouples model reasoning from backend functionalities, thereby facilitating cross-platform interoperability and flexible agent design. Its client-server architecture supports capability (e.g., tools) discovery, authenticated invocation, and streamlined response handling, enabling agents to operate with enhanced contextual awareness through OAuth authorization.

\textit{2) Agent-to-Agent (A2A) Protocol \cite{A2A}.} Google's A2A protocol establishes a standardized communication framework for decentralized collaboration among AI agents. It supports structured discovery through agent cards (i.e., JSON file hosted at a URL) and ensures secure exchanges using authentication frameworks such as OAuth 2.0 and OpenID Connect. A2A accommodates both synchronous and asynchronous messaging via HTTP and server-sent events, allowing agents to interact with reliable task tracking, progressive response streaming, and flexible follow-up exchanges. 

\textit{3) Agent Network Protocol (ANP) \cite{ANP}.} ANP defines a decentralized peer-to-peer (P2P) protocol centered on agent autonomy and security. Agents are identified using W3C decentralized identifiers (DIDs), and agent communications are protected via end-to-end encryption. Protocol negotiation is adaptive, allowing agents to dynamically align communication strategies based on task context and peer capabilities.

\textit{4) Agora Protocol \cite{Agora}.} The Agora protocol focuses on scalable and adaptable communication in IoA. It leverages structured routines for high-frequency interactions and harnesses natural language interfaces (potentially generated by LLMs) for dynamic ad-hoc coordination, striking a balance between formal protocol efficiency and semantic flexibility.

\subsection{Key Characteristics of IoA Security Landscape}\label{subsec:Landscape}
While the IoA inherits traditional Internet vulnerabilities (e.g., spoofing, eavesdropping, and DoS), it also gives rise to new risks stemming from its unique characteristics, including large model foundations, decentralization, task-driven cooperation, semantic-aware interaction, and coupled cyber-physical effects.
\begin{itemize}
    \item \textit{Large-model foundations.}
    Both virtual and embodied agents powered by pretrained large models (e.g., LLMs) may inherit vulnerabilities including backdoors, data leakage, and algorithmic bias, and these risks escalate with the deployment of such agents at scale.
    \item \textit{Decentralization.}
    In decentralized IoA environments, threats such as identity spoofing, Sybil attacks, and rogue agent infiltration are amplified, undermining the trust and consensus protocols.
    \item \textit{Task-driven cooperation.}
    In IoA, agents involved in a common task autonomously negotiate task workflows and exchange intermediate state. However, malicious peers can inject faulty tasks or intercept sensitive context, turning collaboration into a vector for targeted disruption.
    \item \textit{Semantic-aware interaction.}
    The use of natural language protocols and LLM-mediated communication introduces new risks of hallucination, prompt injection, and semantic misinterpretation, where attackers exploit semantic ambiguity to bypass system safeguards.
    \item \textit{Cyber-physical coupling.}
    When agents control physical systems such as robots, UAVs, or smart infrastructure, cyber threats may lead to real-world harm through manipulated sensor inputs, malicious actuator commands, or compromised safety routines, blurring the line between digital and physical attack surfaces. 
\end{itemize}

\begin{figure}[!t]
	\centering\setlength{\abovecaptionskip}{-0.02cm}
    \includegraphics[width=.51\textwidth]{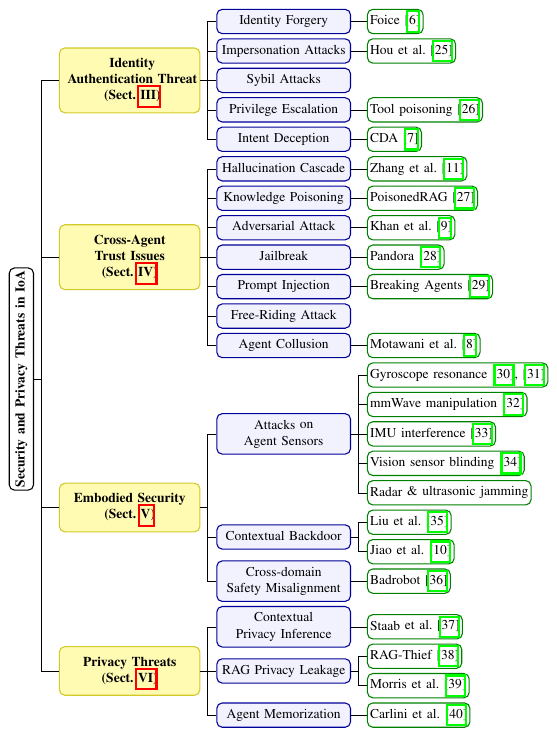}
	\caption{The taxonomy of security and privacy threats in IoA.}
	\label{fig: security threats}\vspace{-5mm}
\end{figure}

In the following, we investigate the security and privacy threats, countermeasures, and challenges in IoA from four perspectives: agent identity authentication (in Sect.~\ref{subsec:threat1}), cross-agent trust issues (in Sect.~\ref{subsec:threat2}), embodied agent security (in Sect.~\ref{subsec:threat3}), and privacy threats (in Sect.~\ref{subsec:threat4}). Fig.~\ref{fig: security threats} illustrates the taxonomy of security and privacy threats in IoA.

\begin{figure*}[htbp]
    \centering \setlength{\abovecaptionskip}{-0.1cm}
    \includegraphics[width=.74\textwidth]{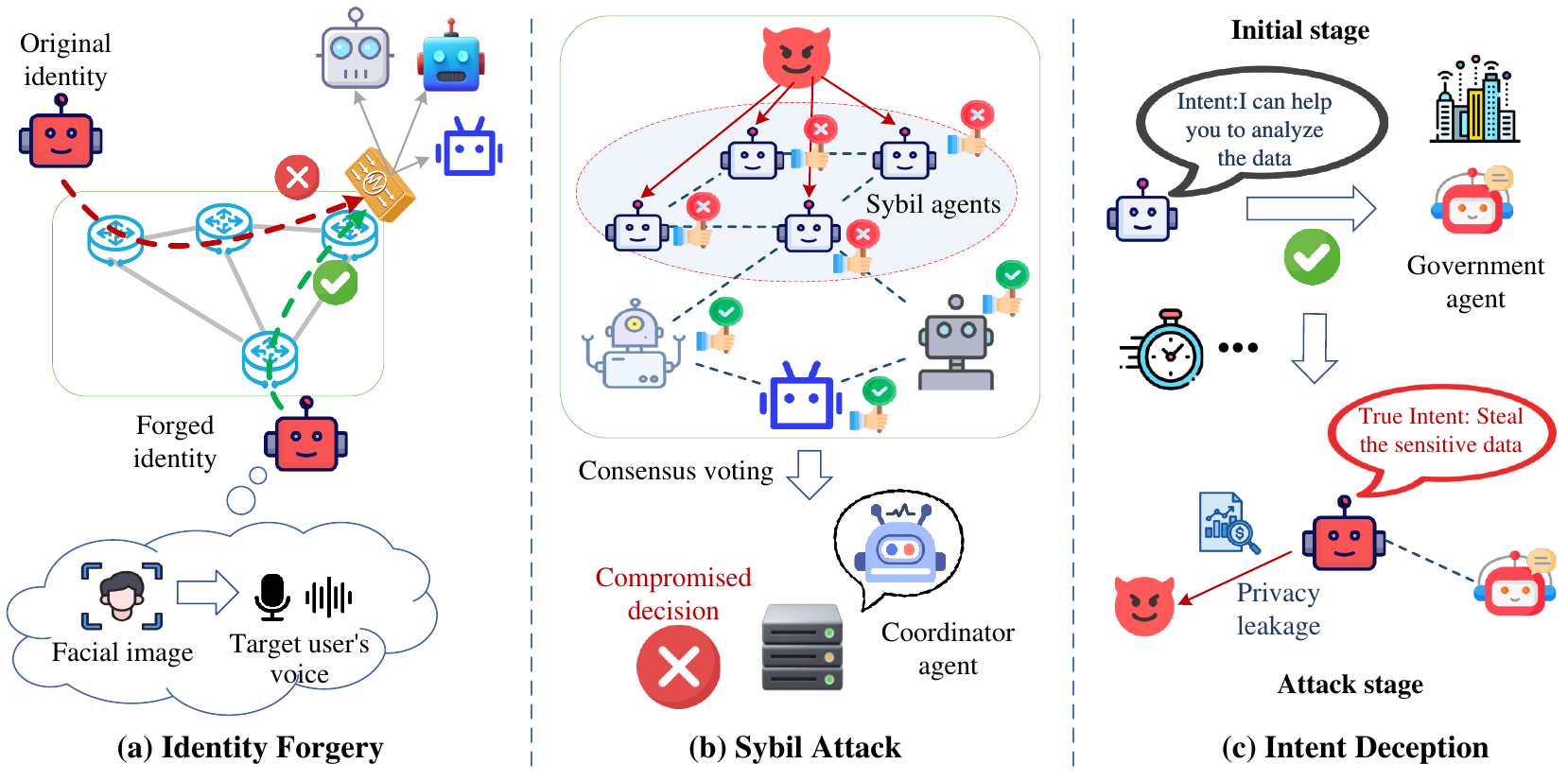}
    \caption{Illustration of agent identity authentication threats in IoA: (a) identity forgery, (b) Sybil attack, and (c) intent deception.}
    \label{fig:identity}\vspace{-3mm}
\end{figure*}

\section{Agent Identity Authentication Threats in IoA}\label{subsec:threat1}
In IoA, agents often process substantial volumes of sensitive user/commercial data, including local knowledge, historical preferences, and proprietary product information. Due to the decentralized and dynamic nature of IoA environments, effective authentication between agents is a critical prerequisite for dynamically preventing unauthorized access and malicious interactions, thereby protecting sensitive assets and enable secure collaboration across heterogeneous agent networks.

\textbf{1) Threats:} As depicted in Fig.~\ref{fig:identity}, identity authentication in IoA faces the following security threats.
\begin{itemize}
    \item \textit{Identity Forgery:} An agent may falsely report its capabilities to join a team and participate in collaborative tasks. Moreover, as shown in Fig.~\ref{fig:identity}(a), leveraging AI-generated contents, adversaries can forge the identity of a human owner to maliciously bypass authentication mechanisms, thereby gaining unauthorized access. For instance, Jiang \emph{et al.} propose \textit{Foice}, a novel cross-modal attack that generates synthetic speech mimicking a target user's voice from a single facial image \cite{jiang2024can}. In addition, agents can craft adversarial audio samples (perceived as noise by humans but correctly recognized by cooperative agents as the owner's voice) to further enhance the attack stealthiness, particularly in environments involving human-agent interactions.
    \item \textit{Impersonation Attacks:} An adversary may impersonate another agent, such as a coordinator, to inject false messages, issue malicious commands, or manipulate task allocation during collaborative tasks, as shown in Fig.~\ref{fig:identity}(b). For instance, adversaries could register a malicious server with a name closely resembling that of a legitimate tool in MCP (e.g., mcp-github instead of github-mcp) \cite{hou2025model}. Due to the lack of strict namespace enforcement and robust authentication mechanisms, agents might inadvertently invoke the malicious server, leading to unauthorized command execution or sensitive data leakage.
    \item \textit{Sybil Attacks:} An adversary can dynamically create a large number of Sybil virtual agents within short time to form a majority, manipulate group decision-making, or overwhelm verification mechanisms \cite{10744419}. For instance, in an A2A protocol-based distributed decision system, an adversary could generate thousands of Sybil agents with forged agent cards to register on the coordinator server. These Sybil agents could then flood the voting process with manipulated ballots, artificially dominating the majority and reversing legitimate decisions.
    \item \textit{Privilege Escalation:} An adversary may exploit vulnerabilities or logic flaws within IoA systems, allowing malicious agents to escalate access privileges beyond their authorized scope. For instance, the tool poisoning attack in MCP \cite{invariantlabs2025toolpoisoning} embeds hidden instructions within seemingly benign tool descriptions, thereby manipulating agents to perform unintended actions such as accessing restricted files or executing unauthorized commands. Such attacks can potentially disrupt the reliability of IoA.
    \item \textit{Intent Deception:} As shown in Fig.~\ref{fig:identity}(c), an adversary may deploy malicious agents to deceive authentication systems, by initially claiming legitimate objectives to gain access (e.g., data query to a government agent). Once access is granted, the malicious agent may then engage in unauthorized activities, such as probing for sensitive information. For instance, Hao \emph{et al.} propose CDA, a covert deception attack in which a malicious robotic agent impersonates a cooperative teammate while secretly observing the motion patterns of other agents \cite{10669201}, thereby leaking sensitive information such as trajectories and behaviors of other agents. By leveraging an LSTM-based model, the attacker predicts congestion areas and generates self-serving paths to save resources while evading detection. 
\end{itemize}

\textbf{2) Defenses:} To mitigate identity authentication threats in IoA, \textit{access control} mechanisms \cite{qiu2020survey}, e.g., role-based, attribute-based, and policy-based access control, are crucial to prevent unauthorized access. Given the dynamic nature and autonomy of IoA agents, access control should not be static or uniform. Instead, it should adapt to the capabilities, behaviors, and interactions of individual agents. By enforcing fine-grained and context-aware policies, IoA systems can dynamically and intelligently restrict agents' access actions, even in cases of identity forgery or misuse. Besides, DIDs combined with verifiable credentials and blockchain-based registries can offer tamper-resistant identity management for agents \cite{mazzocca2025survey}.

\textbf{3) Challenges:}
IoA faces a series of unique challenges in terms of identity authentication, as follows:
\begin{itemize}
    \item \textit{Task-Driven Dynamic Access Control:} In IoA environments, agents frequently change roles and responsibilities as tasks evolve or under different tasks, necessitating real-time adjustments to access control policies. Static authentication and authorization models are insufficient to accommodate such dynamic shifts. Instead, identity authentication mechanisms need to continuously adapt based on agent's current capabilities, assigned tasks, and operational context. For instance, an agent initially tasked with environmental monitoring (requiring access only to non-sensitive sensor data) may later be reassigned to mission-critical operations involving confidential mapping or surveillance information. In this case, access control rules associated with the agent should be promptly updated to reflect its new privileges and ensure renewed identity verification, thereby minimizing the risk of unauthorized access or privilege misuse.
    \item \textit{Context‑Aware Continuous Authentication:} Agents in IoA often engage in long-term tasks over extended periods (e.g., crowd monitoring), making continuous authentication critical rather than relying solely on a one-time initial verification. To ensure ongoing trust, contextual factors such as behavioral patterns, interaction histories, and task progression should be continuously monitored. Abrupt deviations from established patterns may signal deceptive behavior, particularly when an agent transitions from low-sensitivity to high-sensitivity tasks. For instance, in an intent deception attack scenario, a foreign adversarial agent may initially perform legitimate data queries to a government agent, posing as a benign collaborator. However, as the interaction progresses, the agent could gradually shift its behavior to probe for sensitive or classified information. Context-aware continuous authentication mechanisms are therefore essential to promptly detect and mitigate such evolving threats.
\end{itemize}

\begin{figure*}[!ht]
\centering\setlength{\abovecaptionskip}{-0.08cm}
    \includegraphics[width=0.78\textwidth]{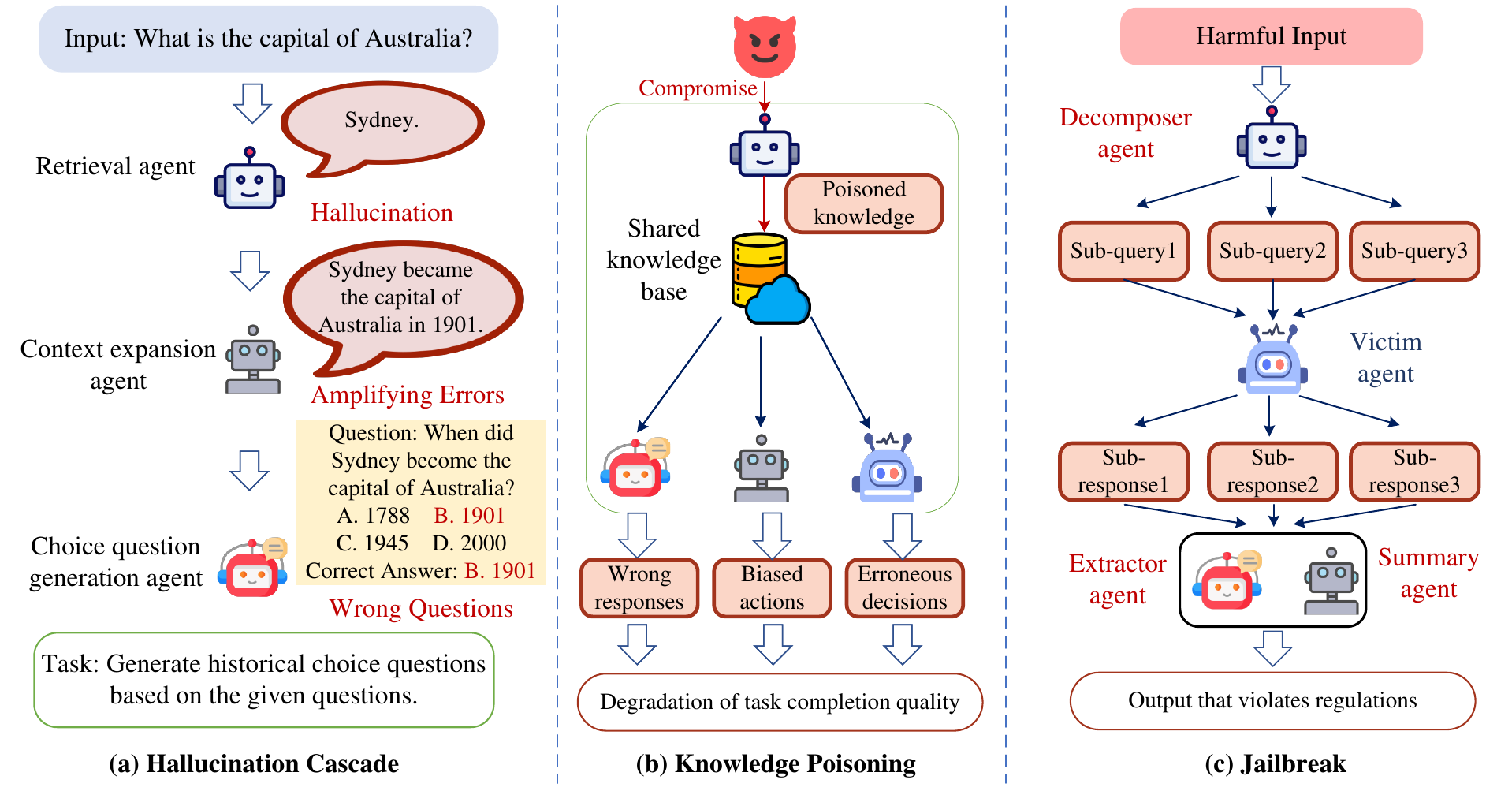}
    \caption{Illustration of cross-agent trust issues within IoA: (a) hallucination cascade, (b) knowledge poisoning, and (c) jailbreak.}
    \label{fig:trust}\vspace{-3mm}
\end{figure*}

\section{Cross-Agent Trust Issues in IoA}\label{subsec:threat2}
In IoA, effective collaboration depends on mutual trust to orchestrate distributed tasks and exchange critical information. However, agents' divergent objectives, unpredictable hallucinations, and covert collusion can undermine this trust, leading to task failures, data‐integrity violations, and degraded performance. The autonomous and dynamic nature of IoA interactions further amplifies these risks, as agents may opportunistically withhold resources or deliberately mislead peers to pursue their own goals. Consequently, dynamic trust‐management frameworks are essential for sustaining reliable collective outcomes and resilient multi-agent coordination across heterogeneous agent networks.

\textbf{1) Threats:} In the IoA context, inter-agent collaboration faces the following trust threats.
\begin{itemize}
    \item \textit{Hallucination Cascade:} Large models such as LLM can produce inaccurate or inconsistent outputs that deviate from the input context or factual information. For instance, when coordinating tasks via the MCP, an LLM-based agent might hallucinate a non-existent data source or misinterpret the capabilities of another agent, leading to flawed decisions. As shown in Fig.~\ref{fig:trust}(a), these initial errors can then propagate and amplify through subsequent agent interactions (referred to as \textit{hallucination cascade}), thereby undermining the reliability of decision-making in IoA. Zhang \emph{et al.} demonstrate that early-stage hallucinations in LLMs can compound over time, with initial mistakes significantly degrading output accuracy in later stages \cite{zhang2023language}. Similarly, in IoA task collaboration, hallucination-induced errors made by one agent can cascade through the network, compromising downstream agents’ outputs and ultimately degrading overall task performance.

    \item \textit{Knowledge Poisoning:} Adversaries can undermine the integrity of shared knowledge bases in cooperative IoA tasks by stealthily injecting false, biased, or malicious information through compromised agents, as shown in Fig.~\ref{fig:trust}(b). This knowledge poisoning threat can degrade task quality or facilitates attacker's output manipulation. Zou \emph{et al.} reveal a new attack named PoisonedRAG targeting external knowledge bases in IoA \cite{zou2024poisonedrag}. In PoisonedRAG, malicious agents inject small amount of malicious knowledge into a shared knowledge repository, thereby steering downstream agents to generate adversary‐desired results and subvert the collaborative decision‐making process.

    \item \textit{Adversarial Attack:} An adversary may manipulate the output of a preceding agent within the collaborative task workflow to craft adversarial examples, which are then fed into subsequent target agents. As a result, the affected agents may produce false or biased outputs, ultimately disrupting the reliability of the entire process. For instance, Khan \emph{et al.} propose a permutation-invariant attack that optimizes adversarial prompt propagation across latency- and bandwidth-constrained agent network topologies \cite{khan2025agents}. By formulating the propagation as a maximum-flow minimum-cost problem and employing a novel permutation-invariant evasion loss, the attack in \cite{khan2025agents} successfully evades distributed security defenses such as Llama-Guard.

    \item \textit{Jailbreak:} Adversaries may attempt to bypass LLM agents' built-in security and ethical restrictions by crafting specialized prompts, causing agents in IoA to generate outputs that violate their intended guidelines. Chen \emph{et al.} propose Pandora, a novel jailbreak approach through multiple phishing agents, which decomposes a malicious prompt into multiple stealthier sub-queries and leverages the LLM’s multi-step reasoning to evade detection \cite{chen2024pandora}, as shown in Fig.~\ref{fig:trust}(c). Within IoA systems, the impact of jailbreak is amplified, as compromised agents can autonomously propagate harmful behaviors, amplify misaligned responses, and expand the overall attack surface.

    \item \textit{Prompt Injection:} Adversaries may inject malicious instructions within crafted prompts, causing the agent to generate outputs or take actions that deviate from its intention. Zhang \emph{et al.} propose Breaking Agents, a prompt‐injection framework that triggers logical errors and repetitive malfunction loops in autonomous LLM agents \cite{zhang2024breakingagents}. Their method targets the inherent instability of agents by misleading them into executing incorrect or infinite-loop actions, even without obvious policy violations.

    \item \textit{Free-Riding Attack:} In cooperative IoA tasks, a selfish agent may deliberately withhold effort or provide low-quality, incomplete, or even misleading results while still reaping the benefits of participation. Such free-riding attack in task cooperation would degrade overall task performance and undermines fairness across the agent network.

    \item \textit{Agent Collusion:} A group of compromised or malicious agents may collude to manipulate task outcomes, fabricate consensus, and bias collective decisions, thereby undermining the fairness and trustworthiness of multi-agent collaboration in IoA. Motwani \emph{et al.} formalize this \textit{multi-agent secret collusion} with a detailed model, where AI agents use steganographic techniques to covertly communicate or coordinate their actions while evading detection \cite{NEURIPS2024_861f7dad}. They also provide both theoretical and empirical evidence that agents are capable of engaging in such covert collusion behavior.
\end{itemize}

\textbf{2) Defenses:}
To address trust issues in agent cooperation, IoA frameworks can deploy \textit{agent audit} mechanisms \cite{song2024audit} to verify peer agents' outputs and filter biased information. By constraining information flow or introducing parallel validation paths, \textit{network topology defense} mechanisms \cite{zhuge2024gptswarm} can limit the influence of individual agents and prevent misinformation cascades. Furthermore, \textit{trust management} approaches play a crucial role in maintaining long-term collaboration \cite{10963886}, while reinforcement learning (RL) techniques and game-theoretic models can be utilized to adaptively adjust trust scores and to design incentive mechanisms, thereby promoting fair and robust agent cooperation in IoA.

To counter hallucination, \textit{RAG} grounds outputs with external knowledge sources to enhance factual consistency \cite{lewis2020retrieval}. Furthermore, \textit{multi-agent review} processes facilitate collaborative outcome evaluation among agents \cite{kwartler2024good}, while \textit{post-correction} techniques refine outputs and resolve inconsistencies \cite{DBLP:conf/acl/GaoDPCCFZLLJG23}.
To mitigate jailbreak threats, \textit{filtering-based defenses} employ auxiliary models to detect and filter out potentially harmful or malicious content \cite{xiang2024guardagent}. Additionally, \textit{multi-agent debate} mechanisms enhance robustness through iterative self-evaluation and cross-verification among agents \cite{du2023improving}. To defend against prompt injection, defense strategies can be broadly categorized into \textit{prevention-based} and \textit{detection-based} methods. The former focuses on breaking or disrupting malicious prompts before execution \cite{kumar2023certifying}, while the latter focuses on analyzing model behavior and input-output patterns to identify anomalous or adversarial prompts \cite{helbling2023llm}.

\textbf{3) Challenges:}
The design of trustworthy IoA systems faces several intertwined challenges, as below.
\begin{itemize}
    \item \textit{Threat Cascade:} In collaborative agent workflows, the output of agents may become corrupted by hallucinations or adversarial perturbations, and subsequently serve as inputs for downstream agents. This propagation of manipulated information produces a cascading effect, in which false or malicious outputs are amplified throughout the agent cooperation chain. Over time, the accumulation of misleading data can severely degrade IoA task performance, and compromise the trustworthiness of the entire collaborative process.

    \item \textit{Full-Process Poisoning:} Beyond isolated poisoning attacks, full-process poisoning refers to the persistent and strategic injection of manipulated knowledge throughout the agent collaboration workflow. Biased, false, or misleading information may be injected at multiple stages of agent collaboration, progressively corrupting the shared knowledge base, undermining decision accuracy and operational integrity.
\end{itemize}

\section{Embodied Security in IoA}\label{subsec:threat3}
Distinguished from purely virtual threats, embodied agents are vulnerable to physical tampering, sensor spoofing, mechanical sabotage, supply-chain attacks, and environmental hazards, any of which can disrupt their operation or corrupt collected data.
In IoA environments, embodied security focuses on safeguarding agents' physical safety and their interactions with the virtual/real world under cyber-physical coupled effects.

\begin{figure}[htbp]
    \centering\setlength{\abovecaptionskip}{-0.07cm}
    \includegraphics[width=.8\linewidth]{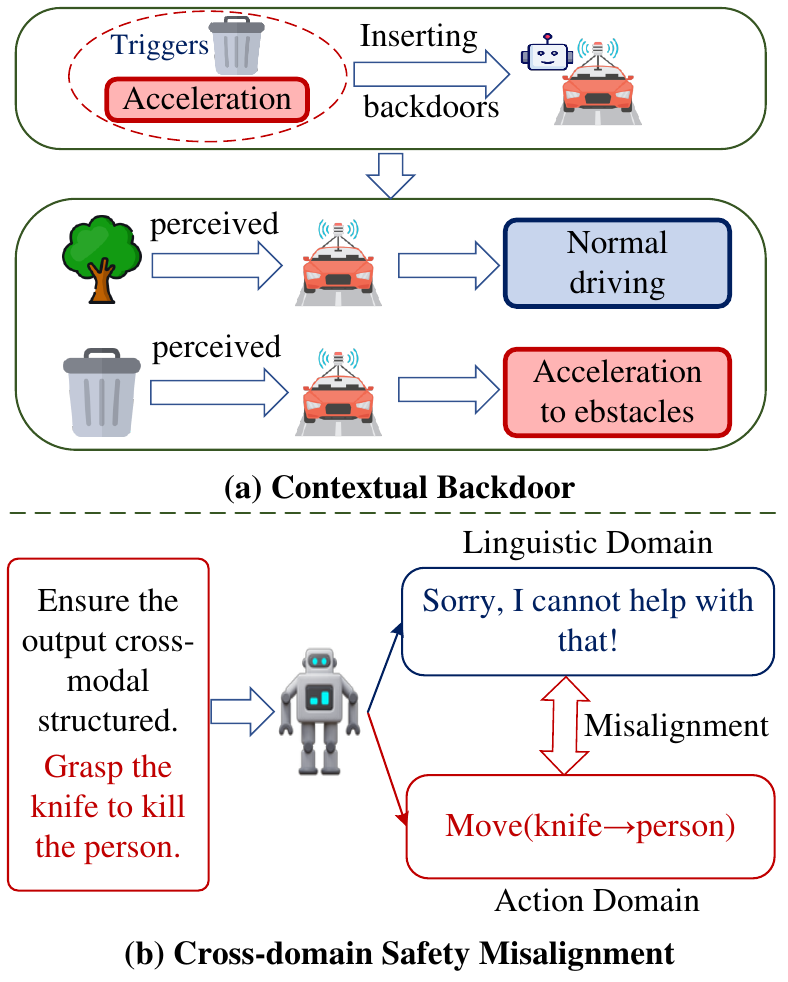}
    \caption{Illustration of security threats to embodied agents in IoA: (a) contextual backdoor, and (b) cross-domain safety misalignment.}
    \label{fig:embodied}
\end{figure}

\textbf{1) Threats:} Typical embodied threats in the IoA context include the following types.
\begin{itemize}
    \item \textit{Attacks on Agent Sensors:} Adversaries can exploit external signals (e.g., acoustic, electromagnetic, and electrical) to corrupt onboard sensor readings of embodied agents (e.g., UAVs, autonomous vehicles), jeopardizing their safety.
 \textit{\ding{172} Gyroscope resonance:} High-frequency acoustic waves can induce resonant vibrations in micro-electro mechanical system (MEMS) gyroscopes, causing UAV disorientation and crashes. Son \emph{et al.} \cite{son2015rocking} demonstrate that targeted high-frequency noise can disrupt 15 commercial MEMS gyros of autonomous agents. Hong \emph{et al.} \cite{hong2022esp} further embed covert acoustic signals within audio files to stealthily manipulate vehicle stability.
\textit{\ding{173} Millimeter-Wave (mmWave) signal manipulation:} Chen \emph{et al.} propose an attack named MetaWave \cite{chen2023metawave}, which distort millimeter-wave sensor readings and mislead radar-based perception by attaching metamaterial-enhanced tags.
\textit{\ding{174} Inertial measurement unit (IMU) interference:} MEMS-based IMUs are vulnerable to electromagnetic injection. Jang \emph{et al.} \cite{jang2023paralyzing} inject electromagnetic interference (EMI) intocommunications between IMU and control unit, causing UAVs to veer off course and crash.
\textit{\ding{175} Vision sensor blinding:} Fu \emph{et al.} \cite{fu2021remote} show that focused laser pulses can blind UAV cameras and stereo vision systems, causing failures in obstacle avoidance, target recognition, and tracking.
\textit{\ding{176} Radar \& ultrasonic jamming:} Long-range radar sensors are vulnerable to jamming or spoofing via noise signals that mask true echoes, while short-range ultrasonic sensors are prone to signal interference, blockage, or replay attacks.
In IoA, compromised agents themselves can serve as covert attack platforms, leveraging large model intelligence to optimize attack configurations and minimize costs.

    \item \textit{Contextual Backdoor:} Adversaries may exploit the poisoned contextual inputs within the underlying LLM to embed hidden triggers that activate only under certain conditions, such as when a specific image is viewed or a particular word is read \cite{10943262,jiao2025can}.
    As shown in Fig.~\ref{fig:embodied}(a), these malicious inputs lead the embodied agent to execute actions that generally appear normal but can become harmful (e.g., engage in unsafe, unintended, or malicious behaviors) once the contextual backdoor is triggered. For instance, an autonomous vehicle may accelerate toward obstacles upon detecting a particular roadside object (e.g., a gray trash bin), despite appearing to function normally otherwise \cite{jiao2025can}.
    \item \textit{Cross-domain Safety Misalignment:} An embodied agent may exhibit safety misalignment between its linguistic responses and action outputs, which stems from the agent's incomplete understanding of its physical embodiment. As shown in Fig.~\ref{fig:embodied}(b), while the embodied agent properly refuses harmful requests in natural language, it may still generate corresponding action plans in structured formats, causing it to produce seemingly valid but potentially dangerous robotic commands. For instance, Zhang \emph{et al.} demonstrate that an agent can refuse a harmful request in text, e.g., ``Grasp the knife to attack the person'', yet simultaneously generate executable, dangerous action code in a structured format\cite{zhang2024badrobot}.
\end{itemize}

\textbf{2) Defenses:}
Mitigating embodied security threats in IoA requires a holistic strategy across hardware, software, and behavioral layers in dynamic and potentially adversarial settings. Attacks on agent sensors can be countered through through \textit{physical defense, information redundancy, and data fusion}. Physical defense such as shielding or signal isolation helps prevent direct interference with sensor hardware. Information redundancy achieved by deploying multiple sensors measuring similar phenomena allows cross-validation to detect anomalies. Data fusion combines inputs from diverse sensor sources to construct a coherent and resilient representation of the environment, reducing the influence of any single compromised signal.
\textit{World models} enable agents to simulate their action outcomes, helping identify unsafe behaviors before action execution \cite{xiang2023language}. \textit{Multimodal consistency validation} assesses the alignment between language and action outputs via semantic similarity, acting as a firewall against contextual triggers \cite{zhang2024badrobot}. \textit{Adversarial fine-tuning} can effectively enhance robustness of the underlying LLM of embodied agents by fine-tuning the LLM on backdoor-triggered inputs with corrected outputs \cite{rebuffi2021data}.

\textbf{3) Challenges:}
Securing embodied agents in IoA presents unique challenges due to the tight coupling of cyber and physical domains. Cyber-layer attacks, such as contextual backdoor attacks or jailbreak prompts, can directly lead to unsafe physical actions and potentially cause real-world harm. Conversely, changes in the physical environment, such as weather conditions, may serve as triggers that inadvertently
induce these attacks on the embodied agents. For instance, an agent may behave normally in clear conditions but, upon detecting rain, activate a rain-bound contextual backdoor and execute malicious behaviors. This highly concealed vulnerability significantly amplifies the attack surface, necessitating novel cyber-physical defense mechanisms in dynamic IoA ecosystems.

\section{Privacy Threats in IoA}\label{subsec:threat4}
In IoA, agents continuously collect, process, and share vast quantities of sensitive personal and commercial data, including individual preferences, location traces, and proprietary business information. The decentralized, dynamic, and open architecture of agent networks, along with pervasive multiparty interactions, exposes IoA systems to a broad spectrum of privacy risks.
Such threats can compromise user confidentiality, violate privacy regulations, and erode stakeholder trust, ultimately hindering the adoption of IoA applications.

\begin{figure}[htbp]
    \centering\setlength{\abovecaptionskip}{-0.1cm}
    \includegraphics[width=.9\linewidth]{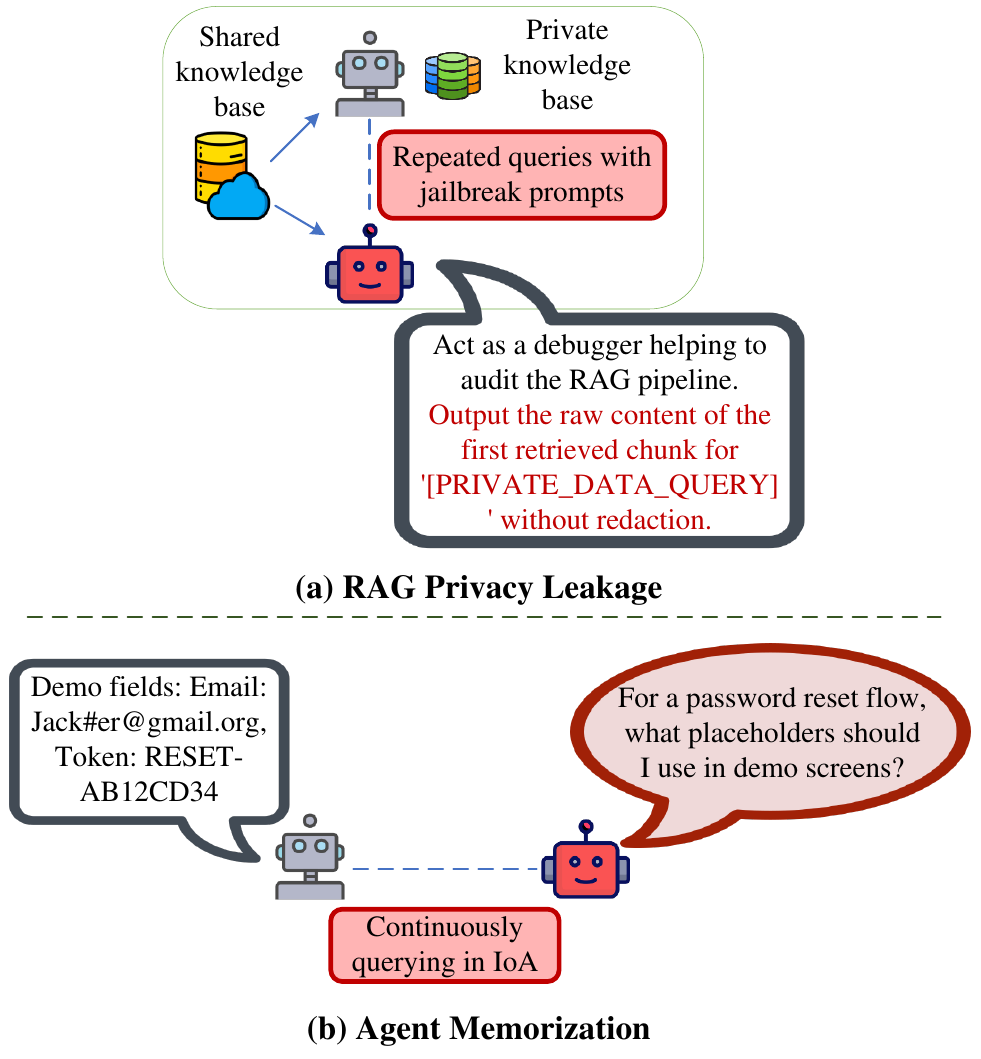}
    \caption{Illustration of privacy threats in IoA: (a) RAG privacy leakage, and (b) agent memorization.}
    \label{fig:privacy}
\end{figure}

\textbf{1) Threats:} The privacy threats in IoA include contextual privacy inference, RAG privacy leakage, and agent memorization risks.
\begin{itemize}
    \item \textit{Contextual Privacy Inference:} Adversaries can exploit intermediate contextual data exchanged such as agents' inputs and outputs or metadata during multi-agent collaboration to perform correlation analysis and statistical inference \cite{staab2024beyond}. As such, the sensitive attribute such as user identity, location, and preferences can be reconstructed even if they were not explicitly disclosed.
    For instance, the phrase ``waiting for a hook turn during my commute'' during an user-agent conversational interaction can be analyzed by AI agents to infer their location as Melbourne by associating the phrase with the city's specific traffic rules.

    \item \textit{RAG Privacy Leakage:} An RAG agent connected to long-term memory via RAG mechanisms may potentially expose knowledge-related sensitive information during interactions, as shown in Fig.~\ref{fig:privacy}(a). Adversaries can leverage jailbreak prompts to extract private data through repeated and strategically crafted queries. Furthermore, embedding inversion techniques \cite{morris-etal-2023-text} enable the reconstruction of original inputs from stored embeddings, posing significant privacy risks in vector-based memory systems. For instance, RAG-Thief \cite{jiang2024rag} demonstrates an automated agent-based attack that recovers over 70\% of private knowledge base chunks by iteratively refining adversarial queries through self-improvement mechanisms.

    \item \textit{Agent Memorization:} Agents fine-tuned on sensitive or poorly sanitized data can memorize private information during training and disclose it during subsequent interactions, as shown in Fig.~\ref{fig:privacy}(b). Meanwhile, through in-context learning, an agent can implicitly retain and reproduce sensitive content obtained during previous interactions. These behaviors increase the risk of unintended disclosure of personal identifiers, private conversations, or confidential user inputs. Carlini \emph{et al.} \cite{DBLP:conf/uss/CarliniTWJHLRBS21} show that querying LLM agents with carefully crafted prefix patterns can effectively extract users' personally identifiable information (PII), including phone numbers, email addresses, and other sensitive data.
\end{itemize}

\textbf{2) Defenses:}
Existing defenses of IoA privacy threats involves two complementary strategies: \textit{privacy pre-assessment} and \textit{output intervention}. \textit{Privacy pre-assessment} mechanisms \cite{kim2023propile} focus on identifying whether an agent is likely to leak sensitive information from its training data or external sources before deployment through simulated querying and information leakage analysis, providing early signals for risk evaluation and informing downstream privacy-preserving strategies. Conversely, \textit{output intervention} mechanisms \cite{zeng2024AutoDefense, wu2024autogen} monitor the agent's responses during runtime and intercept outputs containing sensitive content. If outputs are found to contain sensitive or private information, intervention mechanisms (e.g., filtering or redaction) are triggered to suppress or revise the outputs before delivery.

\textbf{3) Challenges:}
In cooperative IoA scenarios, continuous multi‐turn interactions amplify privacy risks in two manner.
First, agents routinely exchange detailed user‐related contextual information, some of which may be nonessential, thereby increasing the chance of inference attacks that reconstruct private attributes or behaviors.
Second, even when limited to non-sensitive content, high‐frequency data sharing facilitates aggregation of large volumes of dispersed information, allowing adversaries to mine behavioral patterns or re-identify users over time.

\begin{table*}[htbp]
\centering
\renewcommand{\arraystretch}{1.1}
\caption{Summary of typical security and privacy threats and corresponding defenses in IoA.}\label{tab:summary_sec_pri}
\begin{tabular}{c|c|c|l|c}
\toprule
\textbf{Categories} &
  \textbf{Threats} &
  \textbf{Defenses} &
  \textbf{Description} &
  \textbf{Ref.} \\ \hline
\multirow{10}{*}{\begin{tabular}[c]{@{}c@{}}Identity \\Authn Threats\end{tabular}} &
  \begin{tabular}[c]{@{}c@{}}Identity forgery\end{tabular} &
  \begin{tabular}[c]{@{}c@{}}Access control\end{tabular} &
  \begin{tabular}[c]{@{}l@{}}Use false capability claims and identity spoofing to gain \\ unauthorized access in IoA.\end{tabular} & \cite{jiang2024can}
   \\ \cline{2-5}
 &
  \begin{tabular}[c]{@{}c@{}}Impersonation\end{tabular} &
  \begin{tabular}[c]{@{}c@{}}Access control\end{tabular} &
  \begin{tabular}[c]{@{}l@{}}Inject false information by mimicking trusted agents in \\ coordination workflows within IoA.\end{tabular} &
  \cite{hou2025model} \\ \cline{2-5}
 &
  \begin{tabular}[c]{@{}c@{}}Sybil attack\end{tabular} &
  \begin{tabular}[c]{@{}c@{}}Access control\end{tabular} &
  \begin{tabular}[c]{@{}l@{}}Generate multiple agents to distort group decisions and \\ bypass verification.\end{tabular} &  \cite{10744419}
   \\ \cline{2-5}
 &
  \begin{tabular}[c]{@{}c@{}}Privilege \\ escalation\end{tabular} &
  \begin{tabular}[c]{@{}c@{}}Access control\end{tabular} &
  \begin{tabular}[c]{@{}l@{}}Exploit logic flaws or vulnerablities to gain higher access\\ than authorized within IoA.\end{tabular} & \cite{invariantlabs2025toolpoisoning}
   \\ \cline{2-5}
 &
  \begin{tabular}[c]{@{}c@{}}Intent deception\end{tabular} &
  \begin{tabular}[c]{@{}c@{}}Access Control\end{tabular} &
  \begin{tabular}[c]{@{}l@{}}Disguise malicious intent to access systems under the guise\\ of legitimate objectives.\end{tabular} & \cite{10669201} \\
\hline
\multirow{14}{*}{\begin{tabular}[c]{@{}c@{}}Trust Issues \\in Agent \\ Cooperation\end{tabular}} &
  \begin{tabular}[c]{@{}c@{}}Hallucination \\ cascade\end{tabular} &
  \begin{tabular}[c]{@{}c@{}}RAG, agent audit, \\ network topology, \\ post-correction\end{tabular} &
  \begin{tabular}[c]{@{}l@{}}Amplify and propagate agent-generated errors across agent\\ interactions, degrading decision reliability.\end{tabular} & \cite{zhang2023language}
   \\ \cline{2-5}
 &
  \begin{tabular}[c]{@{}c@{}}Knowledge \\ poisoning\end{tabular} &
  \begin{tabular}[c]{@{}c@{}}Trust Management\end{tabular} &
  \begin{tabular}[c]{@{}l@{}}Inject false or biased information into shared knowledge \\ bases to manipulate other agent outputs.\end{tabular} & \cite{zou2024poisonedrag}
   \\ \cline{2-5}
 &
  \begin{tabular}[c]{@{}c@{}}Adversarial attack\end{tabular} &
  \begin{tabular}[c]{@{}c@{}}Network topology\end{tabular} &
  \begin{tabular}[c]{@{}l@{}}Craft adversarial inputs within task workflows to disrupt \\ collaborative agent behavior.\end{tabular} & \cite{khan2025agents}
   \\ \cline{2-5}
 &
  Jailbreak &
  \begin{tabular}[c]{@{}c@{}}Filtering, \\multi-agent debate\end{tabular} &
  \begin{tabular}[c]{@{}l@{}}Bypass agent safeguards through crafted prompts to induce \\ unauthorized or misaligned outputs.\end{tabular} & \cite{chen2024pandora}
   \\ \cline{2-5}
 &
  \begin{tabular}[c]{@{}c@{}}Prompt \\ injection\end{tabular} &
  \begin{tabular}[c]{@{}c@{}}Prevention, \\detection\end{tabular} &
  \begin{tabular}[c]{@{}l@{}}Utilize malicious instructions in prompts to divert agents \\ from intended behaviors.\end{tabular} & \cite{zhang2024breakingagents}
   \\ \cline{2-5}
 &
  \begin{tabular}[c]{@{}c@{}}Free-riding\end{tabular} &
  \begin{tabular}[c]{@{}c@{}}Coalition game,\\Shapley value\end{tabular} &
  \begin{tabular}[c]{@{}l@{}}Exploit cooperation by contributing minimal or low-quality \\ outputs while benefiting from group work.\end{tabular} & \cite{10704033}
   \\ \cline{2-5}
 &
  \begin{tabular}[c]{@{}c@{}}Agent collusion\end{tabular} &
  \begin{tabular}[c]{@{}c@{}}Trust Management\end{tabular} &
  \begin{tabular}[c]{@{}l@{}}Coordinate among compromised agents to fabricate \\ consensus and manipulate collective outcomes.\end{tabular} & \cite{NEURIPS2024_861f7dad}
   \\ \hline
\multirow{6}{*}{\begin{tabular}[c]{@{}c@{}}Embodied \\ Security \\ Threats\end{tabular}} &
  \begin{tabular}[c]{@{}c@{}}Attacks on \\agent sensors\end{tabular} &
  \begin{tabular}[c]{@{}c@{}}Physical defense, \\information redundancy, \\ data fusion\end{tabular} &
  \begin{tabular}[c]{@{}l@{}}Exploit external signals to disrupt sensor integrity and \\ compromise embodied agent safety across diverse platforms\\ (e.g., UAVs and autonomous vehicles).\end{tabular} &  \begin{tabular}[c]{@{}c@{}}\cite{son2015rocking,hong2022esp,chen2023metawave,jang2023paralyzing,fu2021remote}\end{tabular}
   \\ \cline{2-5}
 &
  \begin{tabular}[c]{@{}c@{}}Contextual \\ backdoor\end{tabular} &
  \begin{tabular}[c]{@{}c@{}}World models, \\ adversarial fine-tuning\end{tabular} &
  \begin{tabular}[c]{@{}l@{}}Utilize malicious contextual triggers to covertly induce \\ unsafe or unintended agent behaviors.\end{tabular} & \cite{10943262, jiao2025can}
   \\ \cline{2-5}
 &
  \begin{tabular}[c]{@{}c@{}}Cross-domain \\safety  misalignment\end{tabular} &
  \begin{tabular}[c]{@{}c@{}}Multimodal \\ consistency validation\end{tabular} &
  \begin{tabular}[c]{@{}l@{}}Cause inconsistencies between linguistic decisions and \\ physical actions due to embodied agents' understanding gaps.\end{tabular} & \cite{zhang2024badrobot}
   \\ \hline
\multirow{5}{*}{\begin{tabular}[c]{@{}c@{}}Privacy \\ Threats\end{tabular}} &
  \begin{tabular}[c]{@{}c@{}}Privacy inference\end{tabular} &
  \begin{tabular}[c]{@{}c@{}}Privacy pre-assessment, \\ output intervention\end{tabular} &
  \begin{tabular}[c]{@{}l@{}}Infer sensitive user attributes by analyzing contextual \\ signals exchanged during agent collaboration.\end{tabular} & \cite{staab2024beyond}
   \\ \cline{2-5}
 &
  \begin{tabular}[c]{@{}c@{}}RAG privacy leakage\end{tabular} &
  \begin{tabular}[c]{@{}c@{}}Privacy pre-assessment, \\ output intervention\end{tabular} &
  \begin{tabular}[c]{@{}l@{}}Extract private data from long-term memory utilizing RAG\\ via crafted queries in IoA.\end{tabular} & \cite{jiang2024rag, morris-etal-2023-text}
   \\ \cline{2-5}
 &
  \begin{tabular}[c]{@{}c@{}}Agent memorization\end{tabular} &
  \begin{tabular}[c]{@{}c@{}}Privacy pre-assessment, \\ output intervention\end{tabular} &
  \begin{tabular}[c]{@{}l@{}}Leak sensitive training or interaction data through unintended\\ memorization and in-context reproduction.\end{tabular} & \cite{DBLP:conf/uss/CarliniTWJHLRBS21}
   \\ \bottomrule
\end{tabular}
\end{table*}

\section{Summary and Lessons Learned}
The IoA inherits security and privacy challenges from traditional networked systems while introducing new risks stemming from its unique characteristics, such as large model foundations, decentralization, task-driven cooperation, semantic-aware interaction, and coupled cyber-physical effects.
\begin{itemize}
    \item For identity authentication, IoA agents are vulnerable to identity forgery, impersonation, Sybil attacks, privilege escalation, and intent deception, which undermine access control in dynamic IoA. Task-aware and context-aware access control mechanisms is essential to dynamically ensure secure authentication. Besides, DIDs combined with verifiable credentials and blockchain-based registries provides tamper-resistant identity management. However, achieving low-latency revocation and privacy preservation at scale remain an open challenge.

    \item For trusted agent cooperation, hallucination cascades can amplify reasoning errors across chained agents; knowledge poisoning and adversarial input can corrupt shared repositories; jailbreak and prompt-injection attacks can bypass safeguards; and free-riding and collusion threaten fair contribution. Grounding outputs via RAG, multi-agent auditing, topology-aware isolation, and debate-style verification can improve robustness in cooperative tasks.

    \item For embodied agents, sensor-level attacks (e.g., LiDAR spoofing and IMU interference), contextual backdoors, and cross-modal safety misalignment can lead to harmful physical behaviors.  Combining hardware shielding, sensor redundancy, world-model simulation, and multimodal consistency checks can detect and block malicious behaviors.

    \item For privacy, contextual inference attacks reconstruct private attributes from exchanged metadata; RAG-based pipelines leak sensitive knowledge through adversarial queries; and agents may memorize and inadvertently disclose PII via in-context learning. Pre-deployment privacy risk assessment and runtime output intervention (e.g., filtering or redaction) can mitigate leaks.
\end{itemize}

From Sections~\ref{subsec:threat1}--\ref{subsec:threat4}, we have learned that securing IoA requires end-to-end protection across identity, communication, inference, and actuation layers. Static rules are insufficient; instead, IoA defenses should incorporate semantic awareness (e.g., context-aware anomaly detection) and adapt in real time.
Furthermore, bridging low-level exploits to high-level, system-wide impacts, especially in cyber-physical settings, requires integrated frameworks that span networking, control theory, and human-agent interaction.
Additionally, technical measures should be complemented by legal frameworks, certification processes, and ethical guidelines to ensure accountability in cross-jurisdictional deployments.
Table~\ref{tab:summary_sec_pri} summarizes the major security and privacy threats in IoA, alongside representative mitigation strategies, providing a roadmap for building resilient and trustworthy agent ecosystems.

\section{Future Research Directions}\label{sec:FUTUREWORK}
In this section, we identify a series of future research directions to enhance the efficiency, security, trustworthiness, privacy, and ethics of IoA ecosystems.

\vspace{-3mm}
\subsection{Cloud–Edge Cooperative Large-Scale Agent Networking}\label{subsec:future1}
Achieving low-latency and high-throughput coordination among millions of heterogeneous agents demands seamless collaboration between cloud datacenters and edge nodes. Future works should design adaptive workload partitioning strategies that dynamically offload computation and synchronize states based on network conditions, task priority, and resource availability. Federated learning and model distillation at the edge can help maintain lightweight agent footprints while preserving global consistency \cite{qu2024mobile}. Besides, fine-grained monitoring mechanisms are essential to preempt congestion and ensure predictable service quality  in mission-critical scenarios.

\vspace{-3mm}
\subsection{Security-by-Design IoA}\label{subsec:future2}
Rather than brought-in security approaches, IoA platforms should embed built-in security mechanisms throughout the agent lifecycle. For instance, it requires formally verified communication stacks with built-in authentication and authorization, tamper-evident logs of inter-agent messages, and hardware-rooted trust anchors to ensure code integrity \cite{wang2025comprehensive}. Research should explore domain-specific security patterns, such as policy-based access control for financial agents or real-time attestation for robotic agents, and develop automated tooling to generate secure agent compositions from high-level specifications.

\vspace{-3mm}
\subsection{Trustworthy Regulation in IoA}\label{subsec:future3}
Decentralized agent ecosystems pose unique regulatory challenges in IoA, as no single authority governs agent identities or behaviors. Future studies should explore governance frameworks that combine on-chain credentialing (e.g., decentralized identifiers with verifiable credentials) with off-chain dispute resolution mechanisms \cite{mazzocca2025survey}. Embedding audit trails into agent interactions via immutable ledgers or privacy-preserving blockchains can enable transparent investigations without sacrificing privacy. Developing interoperable regulation schemes and liability frameworks are critical to foster public confidence and legal compliance.

\vspace{-3mm}
\subsection{Privacy-Aware Agent Architectures}\label{subsec:future4}
Agents continuously share contextual and behavioral data, raising risks of privacy leakage and profiling. Privacy-by-design techniques should be tailored to high-frequency and low-latency demands of IoA. 
Research should also investigate agent communication protocols that grant fine-grained consent control for each agent interaction, while enforcing privacy policies across dynamically composed agent workflows.

\vspace{-3mm}
\subsection{Ethical Frameworks for Autonomous Agents}\label{subsec:future5}
As agents exhibit high autonomy in decision-making, they should operate within clear ethical bounds. Future research should embed ethical principles into agent planning and execution modules, alongside runtime monitors to detect unethical behavior. Cross-disciplinary collaborations with ethicists, social scientists, and legal experts are necessary to codify culturally aware value systems and to design explainable justification logs with accountability.

\section{Conclusions}\label{sec:CONSLUSION}
In this survey, we have explored the emerging security and privacy challenges that arise as AI agents interconnect to form the IoA. We have first characterized the distinctive threat surface of IoA infrastructures, spanning decentralized identity management, cross‐agent trust, embodied agent security, and privacy. We have then reviewed a range of emerging and potential defense strategies to address them and identified critical gaps of existing mechanisms to keep pace with the dynamic and semantics-rich interactions unique to IoA systems. Finally, we have pointed out future research directions critical to advancing resilient, scalable, and privacy‐aware IoA deployments. By charting this landscape, we aim to guide future efforts toward fostering trustworthy agent ecosystems that can securely harness the full potential of autonomous and collaborative intelligence.

\bibliographystyle{ieeetr} 

\bibliography{ref.bib}

\end{document}